\documentstyle[12pt,psfig]{article}
\begin{document}
\baselineskip 0.7cm

\newcommand{\gsim}{ \mathop{}_{\textstyle \sim}^{\textstyle >} }
\newcommand{\lsim}{ \mathop{}_{\textstyle \sim}^{\textstyle <} }
\newcommand{\vev}[1]{ \left\langle {#1} \right\rangle }
\newcommand{\lsp}{ \left ( }
\newcommand{\rsp}{ \right ) }
\newcommand{\lmp}{ \left \{ }
\newcommand{\rmp}{ \right \} }
\newcommand{\llp}{ \left [ }
\newcommand{\rlp}{ \right ] }
\newcommand{\labs}{ \left | }
\newcommand{\rabs}{ \right | }
\newcommand{\KEV}{ {\rm keV} }
\newcommand{\MEV}{ {\rm MeV} }
\newcommand{\GEV}{ {\rm GeV} }
\newcommand{\TEV}{ {\rm TeV} }
\newcommand{\mgut}{M_{GUT}}
\newcommand{\mint}{M_{I}}
\newcommand{\mgra}{M_{3/2}}
\newcommand{\mll}{m_{\tilde{l}L}^{2}}
\newcommand{\mdr}{m_{\tilde{d}R}^{2}}
\newcommand{\mllXX}[1]{m_{\tilde{l}L , {#1}}^{2}}
\newcommand{\mdrXX}[1]{m_{\tilde{d}R , {#1}}^{2}}
\newcommand{\mgy}{m_{G1}}
\newcommand{\mgl}{m_{G2}}
\newcommand{\mgc}{m_{G3}}
\newcommand{\nuR}{\nu_{R}}
\newcommand{\slL}{\tilde{l}_{L}}
\newcommand{\slLi}{\tilde{l}_{Li}}
\newcommand{\sdR}{\tilde{d}_{R}}
\newcommand{\sdRi}{\tilde{d}_{Ri}}
\newcommand{\e}{{\rm e}}
\renewcommand{\thefootnote}{\fnsymbol{footnote}}
\setcounter{footnote}{1}

\begin{titlepage}

\begin{flushright}
UT-817
\end{flushright}

\vskip 0.35cm
\begin{center}
{\large \bf  Superheavy Dark Matter with Discrete Gauge Symmetries}
\vskip 1.2cm
K.~Hamaguchi$^{a)}$, Yasunori Nomura$^{a)}$\footnote
{Research Fellow of the Japan Society for the Promotion of Science.}, 
and T.~Yanagida$^{a,b)}$

\vskip 0.4cm

$^{a)}$ {\it Department of Physics, University of Tokyo, 
         Tokyo 113-0033, Japan}\\
$^{b)}$ {\it RESCEU, University of Tokyo, Tokyo 113-0033, Japan}
\vskip 1.5cm

\abstract{We show that there are discrete gauge symmetries protect
 naturally heavy $X$ particles from decaying into the ordinary light
 particles in the supersymmetric standard model.
This makes the proposal very attractive that the superheavy $X$
 particles constitute a part of the dark matter in the present universe.
It is more interesting that there are a class of discrete gauge
 symmetries which naturally accommodate a long-lived unstable $X$
 particle.
We find that in some discrete ${\bf Z}_{10}$ models, for example, a
 superheavy $X$ particle has lifetime 
 $\tau_X \simeq 10^{11}-10^{26}~{\rm years}$ for its mass 
 $M_X \simeq 10^{13}-10^{14}~\GEV$.
This long lifetime is guaranteed by the absence of lower dimensional
 operators (of light particles) couple to the $X$.
We briefly discuss a possible explanation for the recently observed
 ultra-high-energy cosmic ray events by the decay of this unstable $X$
 particle.}

\end{center}
\end{titlepage}

\renewcommand{\thefootnote}{\arabic{footnote}}
\setcounter{footnote}{0}

%
%
%
%

The existence of invisible dark matter in the present universe is one of 
the central issues in particle physics as well as in astrophysics.
Many particle candidates have been, in fact, proposed as constituents of 
the dark matter.
It is usually thought that the dark-matter particles cannot be too
heavy, since stable particles much heavier than $\sim 1~\TEV$ would
easily overclose the universe if they have been once in thermal
equilibrium \cite{GK-PRL}.
Therefore, if one would consider superheavy dark matter, one must
invoke, for example, a late-time entropy production to dilute
sufficiently the abundance of such superheavy relic particles \cite{ELN-PL}.
This may render this scenario unlikely.

Recently, the above problem has been naturally solved by the authors of 
\cite{CKR, KT}.
They have suggested that inflation in the early universe may generate a
desirable amount of such a superheavy particle during the reheating epoch
just after the end of inflation.
Since the production mechanism for the superheavy particle involves only 
gravitational interactions, it is quite independent of detailed
nature of the particle as well as processes of the reheating.
This new observation seems very generic and thus it leads us to
consider, in this paper, the proposal that most of the dark matter in 
the universe indeed consists of some superheavy particles \cite{ELN-PL}.
The numerical calculation in ref.~\cite{CKR} shows that the desired 
abundance of the superheavy particle $X$ is obtained when its mass $M_X$ 
lies in the region $0.04 \lsim M_X / H \lsim 2$ with the Hubble
parameter at the end of inflation $H \sim 10^{13}~\GEV$.
In this paper, we assume $M_X \simeq 10^{13}-10^{14}~\GEV$.

For the $X$ particle to be a dominant component of the dark matter in
the present universe, it must live longer than the age of the universe.
Such stability of the superheavy $X$ particle may be explained by
imposing some symmetries.
We consider here gauge (local) symmetries, since any global symmetries
are broken explicitly by topological effects of gravity \cite{Coleman} 
and the $X$ particle may no longer survive until the present.\footnote{
The lifetime of the $X$ particle can be longer than the age of the
universe if topological effects of gravity are extremely small
\cite{BKV-PRL}.}

There are two classes of gauge symmetries, continuous and discrete.
We concentrate our discussion on discrete gauge symmetries
\cite{discrete_gauge}, since analyses for continuous (non-Abelian) gauge 
symmetries have been addressed in \cite{KR, BEN}.
A discrete gauge ${\bf Z}_2$ symmetry \cite{IR-NP} (called R-parity) 
is already used to avoid the rapid proton decay due to dimension 4
operators in the supersymmetric (SUSY) standard model.
We make our analysis in the framework of SUSY standard model introducing 
one superheavy $X$ particle.
We impose a ${\bf Z}_N$ gauge symmetry in addition to the R-parity and
discuss if there are consistent discrete gauge symmetries that guarantee 
the long lifetime of the $X$ particle required for the dark matter.
We find various examples for such discrete symmetries.
It may be very interesting that some of them naturally accommodate a
superheavy $X$ particle of mass $M_X \simeq 10^{13}-10^{14}~\GEV$ with
long lifetime $\tau_X \simeq 10^{11}-10^{26}~{\rm years}$.
We briefly comment on a possible explanation for the recently observed
ultra-high-energy cosmic ray events by the decay of this unstable
superheavy particle.

First of all, $N$ should be even so that the mass term for $X$ is
allowed in superpotential as
\begin{eqnarray}
  W = \frac{M_X}{2} X^2 .
\end{eqnarray}
Here, we assume that $X$ carries $N/2$ of the ${\bf Z}_N$ charge.
Its R-parity could be even or odd.
${\bf Z}_N$ charges for the SUSY standard-model particles are given by
the following discussion.
\begin{enumerate}
\def\theenumi{\roman{enumi}}
\def\labelenumi{(\theenumi)}
\item All terms in superpotential present in the SUSY standard model
      must be allowed by ${\bf Z}_N$.
\item All anomalies for ${\bf Z}_N$ should vanish (cancellation with 
      the Green-Schwarz term \cite{GS-PL} will be given later).
\end{enumerate}
As for R-parity for the light particles we adopt the charges given
in ref.~\cite{IR-NP}.

The first condition (i) reduces the number of independent ${\bf Z}_N$
charges for the light particles as shown in Table~\ref{discrete_charge}.
Here, we have chosen the ${\bf Z}_N$ charge for a Higgs doublet
$H_u$ to be zero.
This can be done by using a gauge rotation of U(1)$_Y$ in the
standard model without a loss of generality.

From the second requirement (ii) we obtain a set of constraints 
\cite{Ibanez}:
\begin{eqnarray}
  \left\{
    \begin{array}{@{\,}l} 
      0 = \frac{1}{2} r' N \\ \\
      \frac{9}{2}m + \frac{3}{2}p = \frac{1}{2} r'' N \qquad\qquad 
        (r',r'',r''' \in {\bf Z}) \\ \\
      3p = \frac{1}{2} r''' N.
    \end{array}
  \right.
\label{an_cancel_cond}
\end{eqnarray}
Here, $p,m$ are ${\bf Z}_N$ charges for the SUSY standard-model
particles given in Table~\ref{discrete_charge}.
\begin{table}
\begin{center}
\begin{tabular}{|c|cccccccc|}  \hline 
  & $Q$ & $\bar{u}$ & $\bar{d}$ & $l$ & $\bar{e}$ & $H_u$ & $H_d$ 
     & $X$ \\ \hline  
  ${\bf Z}_N$ & $m$ & $-m$ & $-m$ & $p$ & $-p$ & $0$ & $0$ & $N/2$ \\ 
  R & $-$ & $-$ & $-$ & $-$ & $-$ & $+$ & $+$ & $+$ or $-$ \\ \hline
\end{tabular}
\end{center}
\caption{${\bf Z}_N$ discrete charges for the SUSY standard-model
 particles and $X$.
 Here, $p,m = 1,\cdots,N-1$.
 R denotes the R-parity.
 $Q, \bar{u}, \bar{d}, l$ and $\bar{e}$ denote SU(2)$_L$-doublet quark,
 up-type antiquark, down-type antiquark, SU(2)$_L$-doublet lepton and
 antilepton chiral multiplets.
 $H_u$ and $H_d$ are chiral multiplets for Higgs doublets.}
\label{discrete_charge}
\end{table}
The first equation comes from the cancellation of $\{ {\bf Z}_N \} 
\{ {\rm SU}(3)_C \}^2$ anomalies, the second from the cancellation of 
$\{ {\bf Z}_N \} \{ {\rm SU}(2)_L \}^2$ anomalies, and the last from the 
cancellation of ${\bf Z}_N$-gravitational anomalies.
Notice that the anomaly-free condition for discrete gauge symmetries is
much weaker than that for continuous ones \cite{Ibanez}.

These equations (\ref{an_cancel_cond}) are satisfied only if
\begin{eqnarray}
  \left\{
    \begin{array}{@{\,}l} 
      p = r_1 \left( \frac{N}{6} \right) \\
      \qquad\qquad\qquad\qquad\qquad\quad (r_1,r_2 \in {\bf Z}) \\
      m = (2r_2-r_1) \left( \frac{N}{18} \right).
    \end{array}
  \right.
\label{an_free_cond}
\end{eqnarray}
We discard the trivial case, $p=m=0$.
Then, we find nontrivial discrete gauge symmetries,
\begin{eqnarray}
  {\bf Z}_2 &:& (p,m) = (1,1), \label{Z_2}\\
  {\bf Z}_6 &:& (p,m) = (1,1),(2,0), \label{Z_6}\\
  {\bf Z}_{18} &:& (p,m) = (0,4),(3,1),(3,5),
                   (6,2),(6,4),(9,1). \label{Z_18}
\end{eqnarray}
There are other solutions which satisfy eq.~(\ref{an_free_cond}), but
they are trivial embeddings of the above ${\bf Z}_2$, ${\bf Z}_6$ or
${\bf Z}_{18}$ in higher $N$, or charge conjugations or U(1)$_Y$-gauge
equivalents of some of eqs.~(\ref{Z_2}), (\ref{Z_6}) and (\ref{Z_18}).
Notice that if the ${\bf Z}_N$ charge $m$ is even the $p$ is even, and
if the $m$ is odd the $p$ is odd, since $N$ is even (see the second
eq. of (\ref{an_cancel_cond})).
Furthermore, the symmetry (\ref{Z_2}) is nothing but the 
matter parity, which is equivalent to the R-parity in the context of the 
SUSY standard model.
Thus, nontrivial discrete gauge symmetries besides the R-parity 
are only the ${\bf Z}_6$ symmetry with charges in eq.~(\ref{Z_6}) and
${\bf Z}_{18}$ symmetry with charges in eq.~(\ref{Z_18}).\footnote{
If one takes a basis where the ${\bf Z}_N$ charge for $Q$ vanishes, both 
of the ${\bf Z}_6$ and ${\bf Z}_{18}$ become ${\bf Z}_6$.
Therefore, precisely speaking, the anomaly-free discrete symmetry
independent of the U(1)$_Y$ in the standard model is only ${\bf Z}_6$
besides the R-parity.}

Let us now discuss operators ${\cal O}$ and ${\cal O}'$, which consist
only of the SUSY standard-model particles, defined in either super- or
K\"{a}hler potentials as
\begin{eqnarray}
  W = {\cal O}X \quad , \quad K = {\cal O}'X.
\label{decay_interaction}
\end{eqnarray}
The operators ${\cal O}$ and ${\cal O}'$ must carry $N/2$ (mod $N$)
of the ${\bf Z}_N$ charge to make super- and K\"{a}hler potentials $W$
and $K$ invariant under the ${\bf Z}_N$.
If the ${\bf Z}_N$ charge $m$ is even (in this case, the $p$ is always
even), the operators ${\cal O}$ and ${\cal O}'$ have always even 
${\bf Z}_N$ charge, since all the light particles have even charges in
this case.
Thus, any operators ${\cal O}$ and ${\cal O}'$ consist of the light
particles cannot have charge 3 or 9 $(=N/2)$, and hence the $X$ particle 
is completely stable for the even $m$.

If the $m$ is odd (in this case, the $p$ is always odd), we easily see
that the total number of matter multiplets ($Q, \bar{u}, \bar{d}, l$ and
$\bar{e}$ in Table~\ref{discrete_charge}) in the operators ${\cal O}$
and ${\cal O}'$ must be odd so that the ${\cal O}$ and ${\cal O}'$ carry 
3 (9) of the ${\bf Z}_6$ (${\bf Z}_{18}$) charge.
On the other hand, we find from the ${\bf Z}_N$ charge assignment in 
Table~\ref{discrete_charge} that any operators which have odd total
number of matter multiplets carry always odd R-parity.
This proves that the $X$ particle with even R-parity is completely
stable and can survive until the present as a superheavy dark matter.
Notice that the condensation, $\langle H_u \rangle \neq 0$ and $\langle
H_d \rangle \neq 0$, does not break any ${\bf Z}_N$.
It is very surprising that anomaly-free condition (\ref{an_cancel_cond}) 
plays an essential role on deriving the conclusion that the discrete gauge 
symmetries ${\bf Z}_6$ and ${\bf Z}_{18}$ in eqs.~(\ref{Z_6}) and
(\ref{Z_18}) can naturally accommodate a stable $X$ particle as long as
$X$ has even R-parity.\footnote{
For the trivial case, $p=m=0$, the $X$ particle is obviously stable for
any $N$, since all light  particles do not have non-vanishing ${\bf
Z}_N$ charges.}

If the $X$ has odd R-parity and the $m$ is odd, it can couple to the
light particles.
The lowest dimensional operators in the ${\bf Z}_6$ and ${\bf Z}_{18}$
models are given in Table~\ref{lowest_operator}.\footnote{
The dimension of the operator ${\cal O}$ consists of $n$
chiral multiplets is the same as that of the operator ${\cal O}'$
consists of $n-1$ chiral multiplets.}
\begin{table}
\begin{center}
\begin{tabular}{|c|c|c|}  \hline 
  $N$  & $(p, m)$ & the lowest dimensional operator \\ \hline  
  $6$  & $(1, 1)$ & $\bar{u}\bar{d}\bar{d} \quad (\in {\cal O})$ \\ 
  $18$ & $(3, 5)$ & $\bar{d}\bar{d}\bar{d}ll \quad (\in {\cal O})$ \\ 
  $18$ & $(3, 1)$ & $\bar{u}\bar{u}\bar{u}\bar{e}\bar{e} 
                   \quad (\in {\cal O})$ \\ 
  $18$ & $(9, 1)$ & $lH_{u} \quad (\in {\cal O})$ \\ \hline
\end{tabular}
\end{center}
\caption{The lowest dimensional operators which X can couple to in the
 case that $X$ has odd R-parity.
If $X$ has even R-parity, it is completely stable as explained in the
 text.}
\label{lowest_operator}
\end{table}
We find that the lifetime of the $X$ particle is shorter
than $10^{-7}~{\rm sec}$.
The $X$ particle with such a short lifetime, however, is astrophysically 
less interesting.

Let us turn to the case where the anomalies for ${\bf Z}_N$ are
cancelled by the Green-Schwarz term \cite{GS-PL}.
As shown in ref.~\cite{Ibanez}, we have the following constraints:
\begin{eqnarray}
  \left\{
    \begin{array}{@{\,}l} 
      0 = \frac{1}{2} r' N + k \delta_{GS} \\ \\
      \frac{9}{2}m + \frac{3}{2}p = \frac{1}{2} r'' N + k \delta_{GS} 
           \qquad\qquad (r',r'',r''' \in {\bf Z}) \\ \\
      3p = \frac{1}{2} r''' N + 24 \delta_{GS},
    \end{array}
  \right.
\label{an_cancel_cond_GS}
\end{eqnarray}
where $k$ is the Kac-Moody level of the SU(3)$_C$ and SU(2)$_L$ gauge
groups, and $\delta_{GS}$ is a constant \cite{Ibanez}.
Here, we have set the Kac-Moody levels of SU(3)$_C$ and SU(2)$_L$ to
be the same in view of the gauge coupling unification.

The solution of eq.~(\ref{an_cancel_cond_GS}) are
\begin{eqnarray}
  \left\{
    \begin{array}{@{\,}l} 
      p = \left( r_1 + \frac{24}{k}r_3 \right) 
               \left( \frac{N}{6} \right) \\
      \qquad\qquad\qquad\qquad\qquad\qquad\qquad (r_1,r_2,r_3 \in {\bf Z}) \\
      m = \left( 2r_2 - r_1 - \frac{24}{k}r_3 \right) 
               \left( \frac{N}{18} \right),
    \end{array}
  \right.
\label{an_free_cond_GS}
\end{eqnarray}
and we have many independent discrete gauge symmetries
which satisfy eq.~(\ref{an_free_cond_GS}).
We have searched them up to $N = 18$, and in Table~\ref{solution_GS} we
show all solutions, $p$ and $m$, except for those in eqs.~(\ref{Z_2}),
(\ref{Z_6}) and (\ref{Z_18}).
From the first and second eqs. of (\ref{an_cancel_cond_GS}), we find that 
if the ${\bf Z}_N$ charge $m$ is odd the $p$ is odd, and if the $m$ is
even the $p$ is even, since $N$ is even.
\begin{table}
\begin{center}
\begin{tabular}{|c|l|}  \hline 
  $N$  & $(p, m)$ \\ \hline  
  $4$  & $(1, 1)$ \\ 
  $8$  & $(1, 5), (3, 7)$ \\ 
  $10$ & $(1, 3), (2, 6), (3, 9), (4, 2)$ \\ 
  $12$ & $(1, 1), (5, 1)$ \\ 
  $14$ & $(1, 9), (3, 13), (5, 3)$ \\ 
  $16$ & $(1, 5), (3, 15), (5, 9), (7, 3)$ \\ \hline
\end{tabular}
\end{center}
\caption{Sets of solutions, $p$ and $m$, for $N \leq 18$ which satisfy
 the anomaly-free constraints (\ref{an_cancel_cond_GS}) relaxed in the
 presence of the Green-Schwarz term.
 For $N = 6$ and $18$, see eqs.~(\ref{Z_6}) and (\ref{Z_18}) in the text.}
\label{solution_GS}
\end{table}

When $N/2$ is even, the ${\bf Z}_N$ charge $m$ (and $p$) must be odd,
otherwise it becomes a trivial extension of a lower $N$ case.
If the $m$ (and $p$) is odd, the total number of matter multiplets in
the operators ${\cal O}$ and ${\cal O}'$ must be even so that the 
${\cal O}$ and ${\cal O}'$ have even ${\bf Z}_N$ charges.
But, we see from the charge assignment in Table~\ref{discrete_charge} that
any operators which have even total number of matter multiplets carry
always even R-parity.
Thus, the $X$ particle with odd R-parity is stable  when $N/2$ is even.

On the contrary, when $N/2$ is odd, the $m$ (and $p$) can be either odd
or even.
If the $m$ (and $p$) is even, the ${\cal O}$ and ${\cal O}'$ never have
odd $(= N/2)$ ${\bf Z}_N$ charges, since all light particles have even
${\bf Z}_N$ charges in this case.
Thus, the $X$ particle can not couple to any operators ${\cal O}$ and
${\cal O}'$ consist of the light particles and hence it is stable for
the even $m$ (and even $p$).
In the case of $m =$ odd ($p =$ odd), the ${\cal O}$ and ${\cal O}'$
must contain odd total number of matter multiplets, and thus the $X$
particle with even R-parity is stable as shown in the previous 
${\bf Z}_6$ and ${\bf Z}_{18}$ models.

We now see that the relaxed anomaly-free condition
(\ref{an_cancel_cond_GS}) still plays an important role to prove the
stability of the superheavy $X$ particle.
(A crucial point is that the ${\bf Z}_N$ charges $p$ is even, if and
only if the $m$ is even.)

If the $m$ (and $p$) is odd and the $X$ has the R-parity opposite to
that in the stable case, it can decay into the SUSY standard-model
particles.
Even in this case, however, the lifetime of the $X$ particle can be
longer than the age of the universe depending on the choice of $N,p$
and $m$.
In fact, for $N = 10$ (i.e. ${\bf Z}_{10}$)\footnote{
For the ${\bf Z}_{10}$ model to be anomaly free, we must take the
Kac-Moody level $k = 5$ in eq.~(\ref{an_cancel_cond_GS}) for the
standard-model gauge groups SU(3)$_C$ and SU(2)$_L$.} 
we have found no nonrenormalizable operator ${\cal O}$ and ${\cal O}'$
making the $X$ particle to decay up to the eighth order.
In the nineth order we find a possible superpotential
\begin{eqnarray}
  W = \left( \frac{1}{M_*} \right)^7 
      \bar{d}\bar{d}\bar{d}\, ll\, lH_{u}lH_{u}\, X,
\end{eqnarray}
in the ${\bf Z}_{10}$ with $(p, m)=(1, 3)$ and $(3,9)$.
The lifetime of the $X$ particle is of the order of
\begin{eqnarray}
  \tau_X \sim \left( \frac{M_*}{M_X} \right)^{14} \frac{1}{M_X}
         \simeq 10^{11}-10^{26}~{\rm years},
\label{Z_10_tau}
\end{eqnarray}
for the cut-off scale $M_* \simeq 10^{18}~\GEV$ and $M_X \simeq
10^{13}-10^{14}~\GEV$.
This is enough long for $X$ to be a perfect superheavy dark matter.
The lowest dimensional operator and the lifetime of $X$ in each model,
($N, p$ and $m$), are also given for $N \leq 18$ in
Table~\ref{lowest_operator_GS}.
\begin{table}
\begin{center}
\begin{tabular}{|c|c|c|c|c|c|}  \hline 
  $N$  & $(p, m)$  & $k$  & the lowest dimensional operator 
       & R-parity & lifetime \\ \hline  
  $4$  & $(1, 1)$  & $16$ & $lH_{u}lH_{u} \quad (\in {\cal O})$ 
       & $+$ & $10^{-22}-10^{-17}~{\rm sec}$ \\ 
  $8$  & $(1, 5)$  & $32$ & $\bar{d}\bar{d}\bar{d}\, ll\, lH_{u} 
         \quad (\in {\cal O})$ 
       & $+$ & $10^{2}-10^{13}~{\rm sec}$ \\ 
  $8$  & $(3, 7)$  & $32$ & $\bar{d}\bar{d}\bar{d}\, ll\, lH_{u} 
         \quad (\in {\cal O})$ 
       & $+$ & $10^{2}-10^{13}~{\rm sec}$ \\ 
  $10$ & $(1, 3)$  & $5$  & $\bar{d}\bar{d}\bar{d}\, ll\, lH_{u}lH_{u} 
         \quad (\in {\cal O})$ 
       & $-$ & $10^{11}-10^{26}~{\rm years}$ \\ 
  $10$ & $(3, 9)$  & $5$  & $\bar{d}\bar{d}\bar{d}\, ll\, lH_{u}lH_{u}
         \quad (\in {\cal O})$ 
       & $-$ & $10^{11}-10^{26}~{\rm years}$ \\ 
  $12$ & $(1, 1)$  & $16$ & $\bar{u}\bar{u}\bar{d}\, \bar{d}\bar{d}\bar{d} 
         \quad (\in {\cal O})$ 
       & $+$ & $10^{-6}-10^{3}~{\rm sec}$ \\ 
  $12$ & $(5, 1)$  & $16$ & $\bar{u}\bar{u}\bar{d}\, \bar{d}\bar{d}\bar{d}
         \quad (\in {\cal O})$ 
       & $+$ & $10^{-6}-10^{3}~{\rm sec}$ \\ 
  $14$ & $(1, 9)$  & $7$  & $\bar{d}\bar{d}\bar{d}\, ll\, 
         \bar{d}\bar{d}\bar{d}\, ll\, lH_{u} \quad (\in {\cal O})$ 
       & $-$ & $10^{35}-10^{56}~{\rm years}$ \\ 
  $14$ & $(3, 13)$ & $7$  & $\bar{d}\bar{d}\bar{d}\, ll\, 
         \bar{d}\bar{d}\bar{d}\, ll\, lH_{u} \quad (\in {\cal O})$ 
       & $-$ & $10^{35}-10^{56}~{\rm years}$ \\ 
  $14$ & $(5, 3)$  & $7$  & $\bar{d}\bar{d}\bar{d}\, ll\, 
         \bar{d}\bar{d}\bar{d}\, ll\, lH_{u} \quad (\in {\cal O})$ 
       & $-$ & $10^{35}-10^{56}~{\rm years}$ \\ 
  $16$ & $(1, 5)$  & $64$ & $\bar{d}\bar{d}\bar{d}\, ll\, 
         \bar{d}\bar{d}\bar{d}\, ll\, lH_{u}lH_{u} \quad (\in {\cal O})$ 
       & $+$ & $10^{51}-10^{76}~{\rm years}$ \\ 
  $16$ & $(3, 15)$ & $64$ & $\bar{d}\bar{d}\bar{d}\, ll\, 
         \bar{d}\bar{d}\bar{d}\, ll\, lH_{u}lH_{u} \quad (\in {\cal O})$ 
       & $+$ & $10^{51}-10^{76}~{\rm years}$ \\ 
  $16$ & $(5, 9)$  & $64$ & $\bar{d}\bar{d}\bar{d}\, ll\, 
         \bar{d}\bar{d}\bar{d}\, ll\, lH_{u}lH_{u} \quad (\in {\cal O})$ 
       & $+$ & $10^{51}-10^{76}~{\rm years}$ \\ 
  $16$ & $(7, 3)$  & $64$ & $\bar{d}\bar{d}\bar{d}\, ll\, 
         \bar{d}\bar{d}\bar{d}\, ll\, lH_{u}lH_{u} \quad (\in {\cal O})$ 
       & $+$ & $10^{51}-10^{76}~{\rm years}$ \\ \hline
\end{tabular}
\end{center}
\caption{The lowest dimensional operators and the lifetimes of the
 superheavy $X$ particle in the anomaly-free ${\bf Z}_N$ models with 
 $N \leq 18$.
 Here, we have assumed the mass of $X$, $M_X \simeq 10^{13}-10^{14}~\GEV$.
 $k$ denotes the Kac-Moody levels of the SU(3)$_C$ and SU(2)$_L$ gauge
 groups.
 For $N = 6$ and $18$, see Table~\ref{lowest_operator}.}
\label{lowest_operator_GS}
\end{table}

More interestingly, if the $X$ particles are unstable, their decays
might explain the ultra-high-energy cosmic ray events \cite{UHE}
observed recently beyond the Greisen-Zatsepin-Kuzmin (GZK) cut-off
energy \cite{GZK}.
If the $X$ particles are the dark matter in the galaxy, the observed air 
showers may be mostly due to gamma rays and nucleons produced by the $X$ 
decays in the halo of our galaxy and they can easily exceed the GZK
cut-off energy if $M_X \gsim 10^{13}~\GEV$ \cite{BKV-PRL, KR}.
If the $X$ particles form a dominant part of the dark matter, we may fit 
the observed fluxes with the lifetime $10^{15}~{\rm years} \lsim \tau_X
\lsim 10^{22}~{\rm years}$ \cite{BEN, KT, BS}.
We should stress that the $X$ particle with odd R-parity in the ${\bf
Z}_{10}$ models meets nicely this condition (see eq.~(\ref{Z_10_tau})).
It may be very encouraging that these ${\bf Z}_{10}$ models can be
embedded into the grand unified SU(5)$_{GUT}$.\footnote{
By using a U(1)$_Y$ gauge rotation one can make the ${\bf Z}_N$
charges for all light particles to be consistent with SU(5)$_{GUT}$.}

So far we have assumed $X$ to be a singlet of the standard-model gauge
groups.
However, our conclusion can be also applied to the case of $X$ being
${\bf 8}$ of SU(3)$_C$ or ${\bf 3}$ of SU(2)$_L$.
It may be interesting that these particles with masses of the order of
$10^{13}-10^{14}~\GEV$ may resolve the problem of apparent discrepancy 
between the gauge-coupling unification and the string scales
\cite{discrepancy}.
\newpage
%
%
%
\newcommand{\Journal}[4]{{\sl #1} {\bf #2} {(#3)} {#4}}
\newcommand{\PL}{\sl Phys. Lett.}
\newcommand{\PR}{\sl Phys. Rev.}
\newcommand{\PRL}{\sl Phys. Rev. Lett.}
\newcommand{\NP}{\sl Nucl. Phys.}
\newcommand{\ZP}{\sl Z. Phys.}
\newcommand{\PTP}{\sl Prog. Theor. Phys.}
\newcommand{\NC}{\sl Nuovo Cimento}

\end{document}